\documentclass[intlimits,twoside,a4paper]{article} 
\usepackage{amsmath,amssymb}
\usepackage{graphicx}
\usepackage{scalerel}
\usepackage{tikz}
\usepackage{bm}
\usepackage[T2A,T1]{fontenc}
\usepackage[cp1251]{inputenc}
\usepackage{enumitem}

\usepackage{cmpj3}
\usepackage{bm}

\articletype{A part of the Special collection to the 100th anniversary of the birth of Ihor Yukhnovskii}

\issue{2025}{28}{3}{33501}
\doinumber{10.5488/CMP.28.33501}

\title[The influence of dimensional crossover]
{The influence of dimensional crossover on phase transitions and critical phenomena in condensed systems}

\author[O. V. Chalyi, E. V. Zaitseva]{O. V. Chalyi\orcid{0000-0002-5755-3875} \thanks{Corresponding author: \email{avchalyi7@gmail.com}},  E. V. Zaitseva\orcid{0009-0008-4132-4374} }
  \address{Bogomolets National Medical University, 13 Shevchenko Blvd., 01601 Kyiv, Ukraine}

\date{Received April 01, 2025, in final form July 02, 2025}

\begin{document}
\maketitle

\begin{abstract}
This article is aimed at studying the effects of the dimensional crossover (DC) on physical properties of condensed systems near phase transition and critical points. Here we consider the following problems: (1) the theoretical provisions that allow to study the effect of spatial confinement on DC near phase transition and critical points; (2) the study of DC in condensed systems with the Ginzburg number $\mathrm{Gi} <1$, where fluctuation effects are described in different ways at the fluctuation, regular and intermediate (crossover) regions; (3) two types of DC were investigated: (a) a decrease in the linear dimensions $L$ to the values of the correlation length of the order parameter fluctuations leads to the conversion of the dependence on thermodynamic variable into a dependence on linear sizes of 3D systems, as well as (b) a further decrease in linear sizes $L$ the 3D--2D or 3D--1D DC happens depending on slitlike or cylindrical geometry, which is determined by the value of the lower crossover dimensionality $d_{\rm LCD}$; (4) it is proposed to extend the known equalities for critical exponents by using the Mandelbrot formula for fractal dimension $D_f$  as a critical exponent; (5) the influence of 3D--2D DC on  the characteristics of the fine structure of the molecular light scattering (MLS) spectrum is studied.
\end{abstract}

\begin{flushright}
	{ \itshape{\bfseries  This review article is dedicated to 
			
			the 100th anniversary of the birth of
			
			the outstanding theoretical physicist
			
			Igor Yukhnovskii}}
	
\end{flushright}

\section{Introduction}

The influence of various thermodynamic variables, as well as spatial confinement, on critical phenomena and phase transitions were actively studied in many systems of experimental, theoretical and practical interest such as bounded fluids and liquid crystals, low-dimensional magnetic systems, confined soft matter systems, few-layer graphene, carbon nanotubes, porous media, biological membranes, vehicles, synaptic clefts, etc.~\cite{1,2,3,4,5,6,7,8,9,10,11,12,13,14,15}.
     The problem that the authors intend to discuss in this review article, dedicated to the 100th anniversary of the birth of the outstanding theoretical physicist Igor Yukhnovskii, can be formulated as follows: \textit {How the results of 3d bulk systems transfer to the results of 3d bounded, 2D, 1D, 0D systems and vice versa?} Obviously, such transitions, which may be called {\itshape{\bfseries ``dimensional crossover''}}, cannot occur abruptly, but must be associated with a fairly smooth and continuous change in physical properties (for example, such critical parameters as critical exponents of scaling laws or critical values of temperature, density, pressure, etc.). The basis for this statement is the results of theoretical studies of the layer-by-layer ordering and the dimensional crossover (DC), as well as the results of computer-simulation studies~\cite{6,7,8,9,10,11,12,13,14,15}.  When theoretically describing a DC, it is necessary to take into account the following factors: 1) specific effects of spatial limitations, 2) numerical values of the critical exponents in systems with various spatial dimensions $d$ belonging to different universality classes, 3) corresponding experimental results.
     
      Here, we  study the effect of changes in the physical properties of spatially bounded (in other words, finite-size or confined) systems as a result of a DC of the following two types~\cite{12,13,14,15}. {\itshape{\bfseries The 1st type DC-1}} corresponds to the transition from 3-dimensional bulk systems to 3-dimensional confined systems, when the linear size $L$ in the direction of spatial confinement approaches the correlation radius~$\xi$ of fluctuations of the order parameter (for single-component fluids, these are density fluctuations). In this case, the dependence of physical properties on thermodynamic variables for large linear sizes $L \gg \xi$ converts into the dependence of these properties on linear sizes in confined systems with $L \lesssim \xi$. {\itshape{\bfseries The 2nd type DC-2}} corresponds to the case when a further decrease of linear sizes $L$ in confined systems could be treated under certain conditions as a smooth change in linear dimensions (for example, a crossover between 3D and 2D systems in slitlike pores or a crossover between 3D and 1D in cylindrical pores).

To study the DC effects, {\itshape{\bfseries the concept of lower crossover dimension (dimensionality) $D_{\rm LCD}$ }} should be introduced (see the 3rd column in table~\ref{tbl-smp1}). The value of the $D_{\rm LCD}$ for real bounded 3D systems (1st~column) determines the limiting spatial dimension of geometric objects (2nd column), when the linear dimensions of systems in the direction of spatial limitation reach their minimum values. As far as the authors know, the $D_{\rm LCD}$ concept was introduced for the first time in~\cite{9}.

\begin{table}[htb]
	\noindent\caption{Numerical values of the lower crossover dimension $D_{\rm LCD}$ for real 3-dimensional bounded systems. }\vskip3mm\tabcolsep4.2pt
	\label{tbl-smp1}
	\centering\noindent{\footnotesize\begin{tabular}{|p{0.3\linewidth}|c|c|}
			\hline
			\multicolumn{1}{|c}
			{\rule{0pt}{4mm} {\bfseries Real bounded 3D systems}}
			& \multicolumn{1}{|p{0.3\linewidth}|}{{\bfseries Corresponding limiting geometric objects}}
			&\multicolumn{1}{|p{0.15\linewidth}|}{{\bfseries Lower crossover dimension $D_{\rm LCD}$}}
			\\%
			\hline%
			\rule{0pt}{3mm}
			Plane-parallel layer, slitlike pore, membrane, synaptic cleft& Molecular plane& 2 \\%
			\hline
			Cylindrical pore, bar, ionic channel&Molecular line& 1 \\%
			\hline
			Spherical or cubic samples, vesicle& Point (one molecule)&0 \\%
			\hline
	\end{tabular}}
\end{table}

There is another important {\itshape{\bfseries concept of universality classes}} that plays a fundamental role in using the theory of phase transitions and critical phenomena to describe similar phenomena in the living and inanimate nature (see, for example,~\cite{12,13,16,17,18,19,20,21} and references there). According to this concept, the phenomena of different nature belonging to the same universality class are characterized by a similar behavior near the critical points and points (lines) of phase transitions if such common conditions are satisfied for all of them. These basic conditions for bulk and bounded systems are presented in table~\ref{tbl-smp2}.

\begin{table}[htb]
\noindent\caption{Necessary conditions of universality classes for bulk and bounded soft matter systems.}\vskip3mm\tabcolsep4.2pt
\label{tbl-smp2}
\centering\noindent{\footnotesize\begin{tabular}{|p{0.46\linewidth}|p{0.46\linewidth}|}
 \hline
 \multicolumn{1}{|p{0.46\linewidth}|}
{\rule{0pt}{5mm} {\bfseries Basic conditions of universality classes for bulk systems}}
 & \multicolumn{1}{|p{0.46\linewidth}|}{{\bfseries Additional conditions of universality classes for bounded systems}}

\\%
\hline%
\rule{0pt}{5mm}1.	The same spatial dimension. & 5.	The same type of geometry (lower crossover dimension). \\%

2.	The same order parameter dimension. & \\%

3.	The same type of intermolecular interaction (short-range or long-range). & 6. The same type of boundary conditions (hydrophilic, hydrophobic, intermediately wetted). \\%

4.	The same symmetry of fluctuation Ginzburg-Landau Hamiltonian (the fluctuation parts of thermodynamic potentials). & 7.	The same physico-chemical properties under consideration (due to non-universal constants in the conditions of extremes for these properties). \\%
\hline
\end{tabular}}
\end{table}

The results obtained for the 2D--3D crossover are used as biomedical applications for studying the size dependence of the characteristics of molecular light scattering (MLS). In this regard, it should be noted (see, for example, \cite{15,22} and references therein) that the John Strutt (Lord Rayleigh) scattering corresponds to such an electrodynamics problem for which only the contribution from the electric dipole is taken into account in a spherical particle with a radius considerably smaller than the wavelength of the electromagnetic radiation incident on this particle. It means that in the formula for the intensity of the scattered light (for the Umov--Poynting vector), only the coherent dipole scattering of the same frequency as the frequency of the incident radiation (the so-called Rayleigh line) is considered. In contrast to Rayleigh scattering, Gustav Mie scattering studies the coherent scattering by a spherical particle of arbitrary radius compared to the wavelength of incident electromagnetic radiation and arbitrary (in general, complex) refractive index. As a result of solving such an electrodynamic problem, Gustav Mie obtained partial contributions to the Umov--Poynting vector of scattered electromagnetic radiation (in particular, the visible range of wavelengths) from electric and magnetic multipoles of various orders.

In addition, the study of the fine structure of the Rayleigh line, which is also called ``Mandelstam--Brillouin scattering'', is of great interest to experimentators and theoreticans. This scattering is caused by the interaction of incident electromagnetic radiation with natural elastic oscillations that occur in condensed media. In what follows, we  consider classical liquids and aqueous suspensions with tumor cells as such media. The fine structure of the Rayleigh line is manifested in the features of the spectral composition of the scattered radiation, which is characterized by the presence of the following three components: the central one, which coincides in frequency with the frequency of the incident electromagnetic radiation, and two frequency-shifted (incoherent) Mandelstam--Brillouin components symmetrically located with respect to the central component. In section~\ref{sec-5}, we  study the effect of the DC-2 dimensional crossover on the characteristics of the Rayleigh line fine structure, namely: (a) the width of the unshifted central component; (b) the frequency shift of the Mandelstam--Brillouin components with respect to the central component; (c) the Landau--Placzek relation which determines the ratio of the integral intensity of the central component and the integral intensities of the two Mandelstam--Brillouin components. The results obtained from the application of the MLS method can be used for early diagnostics of malignant diseases, since the process of uncontrolled proliferation (tissue growth due to cell division and enlargement) has, as it is shown, a significant effect on the size dependences of the fine structure characteristics of the Rayleigh line.

The structure of this review article is as follows. Section~\ref{sec-2}  discusses the basic theoretical background for describing the DC effects on phase transitions and critical phenomena. Section~\ref{sec-3} is devoted to the study DC-1 in condensed systems, in particular with the Ginzburg number $\mathrm{Gi} <1$. In section~\ref{sec-4}, the effects of the DC-2 on the dependence of the critical exponents are studied at 3D--2D DC (see, e.g.,~\cite{25,26,27,28,29,30,31,32,33,34,35,36,37,38}). In addition, the consequences of including the Mandelbrot's formula for the fractal dimension $D_f$ are studied in section~\ref{sec-4} (see, e.g.,~\cite{39,40,41,42,43}). Section~\ref{sec-5} is devoted to the study of the effect of 3D--2D DC on the characteristics of the fine structure of the MLS spectrum and its corresponding biomedical applications.

\section{Theoretical foundations for the study of the effects of spatial limitation}
\label{sec-2}
In this section, we  consider the main theoretical concepts that allow us to consistently study the effect of spatial confinement on the dimensional crossover during phase transitions and critical phenomena in condensed systems using the example of confined liquids.

 First of all, we consider the methods for obtaining correlation functions of the order parameter and the correlation radius, formulas for thermodynamic variables (temperature, density, pressure) that  take into account the effect of spatial confinement, as well as those inequalities that provide the dimensional crossover DC-1. To study the properties of confined fluids we used a method of pair correlation function~(CF) $G_2(r)$ as a Green function of the Helmholtz operator, corresponding to the differential Ornstein--Zernike (OZ) equation with appropriate boundary conditions at the limiting surfaces~\cite{23,24,25,26,27}. The differential OZ equation is derived from the exact integral OZ equation for short-range direct correlation functions (DCF) $C(r)$ considering any number of its spatial moments $C_i$ and short-range intermolecular potentials. Taking into account only the main contributions to CF $G_2(r)$ and hydrophobic (zero) boundary conditions, the following formulae were obtained for $G_2(r)$ in spatially confined systems with the geometry of slitlike and cylindrical pores: 
\begin{equation} 
G_2(\rho^*, z)=(\piup H)^{-1}K_0[\rho^*(\kappa^2+\piup^2/H^2)^{1/2}]\cos(\piup z/H),
\label{eq-1}
\end{equation} 
\begin{equation} 
G_2(\rho^*, z)=C J_0(\mu_1\rho^*/a_0) \exp[-(\kappa^2+{\mu_1}^2/{a_0}^2)^{1/2}|z|].
\label{eq-2}
\end{equation}
Here, $K_0(u)$, $J_0(u)$ are the Macdonald and Bessel functions, correspondingly; $\rho^*=(x^2+y^2)^{1/2}$; $z$ is the coordinate  perpendicular to the bounding surfaces in a slitlike pore or directed along the axis of a cylindrical pore; $\kappa=\xi^{-1}$ is the inverse correlation length $\xi$; $\mu_1 \approx 2.405$ is the first zero of the Bessel function, $a_0$ is the radius of the cylindrical pore; $H$ is the thickness of the slitlike pore; $C$ is a constant coefficient.

Since CF $G_2(r)$ for confined systems do not have an exponential form, it is natural to define the correlation length $\xi$ of the order-parameter fluctuations as a normalized second spatial moment $M_2$ in accordance with the following formula:
\begin{equation}
\xi=\sqrt{M_2}=\sqrt{\frac{\int G_2(r)\, r^2 \,\rd r}{\int G_2 (r)\,\rd r}}. 
\label{eq-3}
\end{equation}

{\itshape{\bfseries Formulas for thermodynamic variables taking into account the effect of spatial confinement.}} Using the approach described above, it becomes possible to calculate the dependence of the physical properties, including the correlation length (radius) of the order parameter fluctuations, not only on the thermodynamic variables but also on linear sizes of confined liquid volumes. For this purpose, the following new temperature, density, and pressure variables should be introduced for liquids in confined geometry, depending on the relationship between the characteristic linear size and the correlation length: 
\begin{equation}
\tau(S, \xi)=(G/S)^{1/\nu}+[1+(G/S)^{1/\nu}](\xi_0/\xi)^{1/\nu} ,      \label{eq-4}
\end{equation}
\begin{equation}
\Delta \rho(S, \xi)=(G/S)^{\beta/\nu}+[1+(G/S)^{\beta/\nu}](\xi_0/\xi)^{\beta/\nu} ,  \label{eq-5}
\end{equation}
\begin{equation}
	\Delta p(S, \xi)=(G/S)^{\beta\delta/\nu}+[1+(G/S)^{\beta\delta/\nu}](\xi_0/\xi)^{\beta\delta/\nu} . 
\label{eq-6}
\end{equation}
Here, the geometrical factor $G=\piup$  for plane-parallel layers, $G=\mu_1=2.4048$   for cylindrical pores, the size parameter $S=L / d_0$ specifies the number of monomolecular layers, where $d_0$  is the diameter of the molecule and $L$ is the linear dimension of the system in the direction of spatial limitation~\cite{26}.

{\itshape{\bfseries Inequalities characterizing the dimensional crossover DC-1.}} Let us consider the important question of obtaining inequalities characterizing the dimensional crossover of the 1st type, i.e., DC-1, when passing from bulk to confined condensed systems.
As it is seen from the formula~\eqref{eq-4}, for large sizes $L \gg \xi$   due to an inequality                                                                                                              
\begin{equation}\xi_0/\xi \gg (G/S)[1+(S/G)^{1/\nu}]^\nu, 
	\label{eq-7}
	\end{equation}
all terms with $(S/G)$  may be omitted and the correlation length $\xi$  is approaching its bulk value $\xi=\xi_0\tau^{-\nu}$. In this case, all physical properties depend on the temperature and on other thermodynamic variables.                                                                                           
In the opposite case of small sizes $L <\xi$  , when such an inequality realizes 
\begin{equation}S=L/d_0<G(\xi/\xi_0)[1+(G/S)^{1/\nu} ]^{-\nu}, 
	\label{eq-8}
	\end{equation}
all physical properties depend on the size parameter $S$ in confined liquids.                                                            
Let us consider, for instance, the case of relatively small linear sizes $L$ with the number of molecular layers $S=L/d_0\approx 10$  and relatively large correlation lengths $\xi_0/\xi \approx 10^{-2}$  for the temperature variable  $\tau \approx 10^{-3}$. In this case, the first term in the parentheses~\eqref{eq-7} becomes 10 times larger than the second term. As a result, all physical properties depend only on the size variable $S$ in such a confined liquid.

\section{DC-1 phenomena in the critical region of condensed matter systems}
\label{sec-3}

{\itshape{\bfseries The Ginzburg number and crossover phenomena for systems with \mathversion{bold}$\mathrm{Gi} < 1$.}} It is known~\cite{16, 32, 33} that the Ginzburg number $\mathrm{Gi}$, which characterizes the fluctuation effects in the vicinity of critical and phase transition points, is determined by the ratio of the mean square fluctuation of the order parameter $\langle \Delta \varphi^2 \rangle$ to the square of the order parameter equilibrium value $ {\varphi_0}^2 $ in accordance with the following formula: 
$$\mathrm{Gi} \sim \langle \Delta \varphi^2 \rangle / {\varphi_0}^2 \sim {\varphi_0}^{d-4} / {\xi_0}^2.$$
It follows from this formula that the fluctuations of the order parameter can be neglected, and the Landau mean-field theory turns out to be valid, if only the value of the spatial dimension $d \geqslant 4$ and (or) the amplitude of the correlation length $\xi_0 \rightarrow \infty$, which is of the order of the radius of intermolecular interactions. Exactly these important conditions  gave Wilson and Fisher~\cite{6, 20} the basis to create an effective method of $\varepsilon$-expansions for calculating the numerical values of critical exponents where $\varepsilon=4-d$ is the deviation of the spatial dimension $d$ from its value in 4-dimensional space. 

There are liquids and other condensed systems for which the Ginzburg number satisfies the inequality $\mathrm{Gi} <1$. A similar situation takes place for the most common as well as the most mysterious liquid, which is water with the Ginzburg number $\mathrm{Gi} \approx 0.3$  (see, for example, the review~\cite{28}). It means that for water and other liquids with a relatively small Ginzburg number $\mathrm{Gi} <1$, the whole critical region, for which the thermodynamic variables, such as deviations of temperature  $\tau=(T_c - T)/T_c$, density $\Delta \rho = (\rho_c-\rho)/\rho_c$ and pressure $\Delta P=(P_c-P)/P_c$  from their critical values, satisfy the inequalities $0<\tau \leqslant 1$, $0<\Delta \rho \leqslant 1$,  $0<\Delta P \leqslant 1$,   can be divided into the following three regions~\cite{14}:

\begin{enumerate}[label=(\alph*)]
	\item {\itshape{\bfseries  the fluctuation region }}
		\begin{equation}
		0<|\tau| \ll \mathrm{Gi}, \quad 0<|\Delta \rho| \ll \mathrm{Gi}^{1/\beta}, \quad 0<|\Delta P| \ll  \mathrm{Gi}^{1/\beta \delta},
		\label{eq-9}
	\end{equation}
	where the fluctuation effects play a decisive role;
	
	\item {\itshape{\bfseries the crossover region  }}
	\begin{equation}
		|\tau| \approx \mathrm{Gi},\quad |\Delta \rho| \approx \mathrm{Gi}^{1/\beta}, \quad |\Delta P| \approx  \mathrm{Gi}^{1/\beta \delta},  
		\label{eq-10}
	\end{equation}					 
	where the fluctuation effects are of the same importance as their background (regular) contributions far from the critical point;
	
	\item {\itshape{\bfseries the regular region}}
	\begin{equation}
		\mathrm{Gi}<\tau \leqslant 1,\quad \mathrm{Gi}^{1/\beta}<\Delta \rho \leqslant 1, \quad \mathrm{Gi}^{1/\beta \delta}<\Delta P \leqslant 1, 
		\label{eq-11}
	\end{equation} 				
	where the fluctuation effects do not play an important role  compared to the background contributions.
	
\end{enumerate}

 In this last region, the Landau mean-field theory is valid, and critical exponents assume their ``classical'' values:
\begin{equation}
	\beta=\nu=1/2, \quad \gamma = 1, \quad \delta=3.
	\label{eq-12}
\end{equation} 
As is known from the fluctuation theory of critical phenomena and the 2nd order phase transition, the Landau mean-field theory may be used for the spatial dimensionality $d=4$, for which the fluctuation effects can be neglected with a logarithmic accuracy. It should be noted here that instead of the words ``the 2nd order phase transition'' it is better to use the words ``continuous phase transitions'', because most of the critical exponents are  fractional numbers rather than integers, as Paul Ehrenfest~\cite{29} once believed.  It means that a transition between the crossover-region behavior and the regular-region behavior can be treated as a 3D--4D DC-1 phenomenon. Thus, for liquids like water such a DC-1 transition takes place in the intervals of temperature $0.30<|\tau| \leqslant 1$, density $0.025<|\Delta \rho| \leqslant 1$  and pressure   $0.45<|\Delta P| \leqslant 1$ .                                                          

{\itshape{\bfseries Results of DC-1 calculations for the diffusion coefficient in a binary liquid mixture and in confined supercooled water (CSW).}} As a DC-1 example in {\itshape{\bfseries the dynamic fluctuation region}}, let us consider such a kinetic (dynamic) physical property as the diffusion coefficient $D$ of a binary liquid mixture, which is determined by the product of the singular part of the Onsager kinetic coefficient $a_s$ and inverse  value of the isobaric-isothermal compressibility $(\partial \mu/\partial x)_{p,T}$ in accordance with the following formula:
\begin{align}
D=&~a_s (\partial \mu/\partial x)_{p,T} \sim L^{1-\gamma/\nu}f_D^{(1)}(y,z) \sim \tau^{\gamma-\nu}f_D^{(2)}(y,z) \sim \Delta x^{(\gamma-\nu)/\beta}f_D^{(3)}(y,z) \nonumber\\  
&\sim \Delta p(S,\xi^*)^{(\gamma-\nu)/\beta\delta}f_D^{(4)}(y,z).      	\label{eq-13}
\end{align}                                                      
Here, $f_D^{(1)}$, $f_D^{(2)}$, $f_D^{(3)}$, $f_D^{(4)}$ are scaling functions for size, temperature, concentration, pressure dependencies, correspondently.

In the dynamic fluctuation region, the following inequalities are fulfilled for thermodynamics variables: 
 $$0\leqslant|\tau|<\tau_D,\quad 0\leqslant|\Delta p|<\Delta p_D,\quad 0\leqslant|\Delta x|<\Delta x_D.$$  
Here $\tau_D=(T_D - T_c)/T_c$, $\Delta P_D=(P_D-P_c)/P_c$,  $\Delta x_D=(x_D-x_c)/x_c$  are the so-called the crossover temperature~\cite{30}, pressure and concentration~\cite{31}, for which $a_s\approx a_r$. In such an immediate vicinity of the phase transition (critical) points, the role of fluctuation effects becomes decisive due to the Ginzburg--Levanyuk criterion~\cite{24, 32, 33}, which can be written in the case under consideration as the following inequalities for thermodynamic variables $\tau$, $\Delta p$, $\Delta x$  and the Ginzburg  number $\mathrm{Gi}$: 
\begin{equation}
0\leqslant|\tau| \ll \mathrm{Gi}, \quad  0 \leqslant |\Delta P| \ll  \mathrm{Gi}^{1/\beta \delta},\quad 0\leqslant |\Delta x| \ll \mathrm{Gi}^{1/\beta}. 
\label{eq-14}
\end{equation} 	                                     
A direct consequence of formula~\eqref{eq-13} is the fact that the diffusion coefficient $D$ takes on a zero value in the most critical state of a binary mixture in accordance with the following power-law dependences: 
\begin{equation}
	D \sim L^{-0.963} \sim |\tau|^{0.607} \sim |\Delta x|^{1.859} \sim |\Delta P| ^{0.388}. 
	\label{eq-15}
\end{equation} 	                                                 
At the same time, it is quite obvious that zero values of such physical properties as the diffusion coefficient, the speed of sound and other quantities at the phase transition (or critical) point contradict the physical considerations as well as, incidentally, infinite large values at the phase transition (or critical) points of the physical properties such as the isothermal compressibility, the isobaric and isochoric heat capacities and other anomalously growing quantities. The obvious reason for such non-physical results is the neglect of the effects of spatial, temporal or spatio-temporal dispersion, i.e., the effects of spatial and temporal non-locality (memory) for physical properties. As was shown, for example, in~\cite{34}, the terms describing the indicated dispersion effects should be added to (a) quantities that tend to zero or to (b) the reciprocal values of physical quantities that tend to infinity at phase transition (or critical) points.
	
The diffusion coefficient of 2D fluids like the confined supercooled water (CSW) belonging to the Ising-model universality class~\cite{12} may have the following power-law critical behavior with taking into account the following values of the critical exponents  $\alpha = 0$,  $\beta= 0.125$,  $\gamma =1.75$,   $\delta= 15$, $\nu  = 1$:
\begin{equation}
	D \sim L^{-0.75} \sim |\tau|^{0.75} \sim |\Delta x|^{6.0} \sim |\Delta P| ^{0.4}. 
	\label{eq-16}
\end{equation} 	                                                  
Concluding the consideration of the physical properties of the bounded liquids in the dynamic fluctuation region, it should be emphasized that it is difficult to conduct experimental studies in this region due to the small value of the dynamic crossover temperature~\cite{15,30}.

In {\itshape{\bfseries the dynamic crossover region}}, where the following inequalities are valid: 
\begin{equation}
\tau_D<|\tau| < \mathrm{Gi}, \quad \Delta p_D < |\Delta P| <  \mathrm{Gi}^{1/\beta \delta}, \quad \Delta x_D< |\Delta x| < \mathrm{Gi}^{1/\beta}, 
	\label{eq-17}
\end{equation} 	                         
and the singular $a_s$ and regular $a_r$ parts of the Onsager kinetic coefficients turn out to be the values of the same order, which allows us to write down the diffusion coefficient in the form
\begin{equation}
D\approx 2 a_r (\partial \mu/\partial x)_{p,T} \sim  L^{-\gamma/\nu} \sim \tau^{\gamma} \sim \Delta x^{\gamma /\beta} \sim  \Delta p^{\gamma/\beta\delta}.
	\label{eq-18}
\end{equation} 	          
For such liquids as water with the Ginzburg number $\mathrm{Gi} < 1$, an important conclusion follows regarding the possibility of conducting experimental studies in the dynamic crossover temperature range $10^{-5} <| \tau| < 0.3$ considered here. Thus, the following power laws for the diffusion coefficient are valid:
\begin{enumerate}[label=(\alph*)]
	\item in 3D bulk liquids
	\begin{equation}
		D\approx 2 a_r (\partial \mu/\partial x) \sim  L^{-1.963} \sim \tau^{1.237} \sim \Delta x^{3.789} \sim  \Delta p^{0.791},
		\label{eq-19}
	\end{equation} 	
	
	\item in 2D bounded liquids like CSW
	\begin{equation}
		D \sim  L^{-1.75} \sim \tau^{1.75} \sim \Delta x^{14} \sim  \Delta p^{0.933}. 
		\label{eq-20}
	\end{equation} 	
\end{enumerate}

Here, a large number in the exponent of concentration dependence $D \sim \Delta x^{14}$ might be explained by the fact that numerical values of the ratio $\gamma / \beta$ in~\eqref{eq-18} are equal to $1.75/0.125 = 14$ for 2-dimensional systems (see table~\ref{tbl-smp3}).
	
In {\itshape{\bfseries the dynamic regular region}}, where such inequalities are valid
\begin{equation}
\mathrm{Gi}<\tau \leqslant 1,\quad    \mathrm{Gi}^{1/\beta \delta}<\Delta p \leqslant 1, \quad \mathrm{Gi}^{1/\beta}<\Delta x \leqslant 1, 
	\label{eq-21}
\end{equation} 	   
                    
\noindent
fluctuation effects do not play a decisive role. Therefore, the critical exponents of the Landau theory $\beta=\nu=1/2$, $\gamma = 1$, $\delta=3$ may be used for obtaining the power laws for the water diffusion coefficient
\begin{equation}
 D \sim  L^{-2} \sim |\tau| \sim |\Delta x|^{2} \sim  |\Delta p|^{2/3}    	\label{eq-22}
 \end{equation} 	                                                         %
in the intervals of temperature  $0.30<|\tau| \leqslant 1,$  pressure  $0.45<|\Delta p| \leqslant 1,$  and concentration  $0.09\ll|\Delta x| \leqslant 1.$
	
Since the Landau theory of phase transition is valid for the spatial dimensionality $d=4$, the crossover transition between dynamic crossover and dynamic regular regions may be considered as a 3D--4D crossover phenomenon. This means that such crossover transitions are realized for the diffusion coefficient~$D$ in the following dependences: on size from  $D\sim L^{-1.963}$  to $D \sim L^{-2},$  on temperature from $D \sim (T - T_c )^{1.237}$ to  $D \sim (T-T_c ),$  on concentration from  $D \sim (x-x_c)^{3.789}$  to $D \sim (x-x_c)^2,$  and on pressure $D \sim (p-p_c )^{0.791}$ to $D \sim (p-p_c )^{0.667}$. 

	Such a transition for CSW between 2D crossover and 4D regular regions should be accompanied by the following changes in diffusion coefficient dependences on: (a)~the size from $D\sim L^{-1.75}$  to $D \sim L^{-2}$, (b)~the temperature from $D \sim (T - T_c )^{1.75}$ to  $D \sim (T-T_c )$, (c)~the concentration from  $D \sim (x-x_c)^{14}$  to $D \sim (x-x_c)^2$,   (d)~the pressure variable from $D \sim (p-p_c )^{0.933}$ to $D \sim (p-p_c )^{0.667}.$ 

 	It is very important to emphasize here that the temperature dependence of the diffusion coefficient in CSW $D \sim (T - T_c )^{1.75}$ has already been confirmed experimentally in~\cite{15}. Therefore, a further experimental verification of the remaining theoretical predictions presented by us is of genuine scientific interest.

1. In reduced geometry, if inequality $\xi>L$ (small volumes) is valid, the 1st term $(G/S)^{1/\nu}$ will prevail in accordance with formula~\eqref{eq-4}. This is the reason why the diffusion coefficient $D$ is decreasing at the fixed temperature with increasing $L$ of the liquid volume.

2. For relatively large sizes $L\gg\xi$, the second term will play a greater role. That is why $D$ will increase,  asymptotically approaching its value $D_0$ for the bulk liquid volume.

The temperature variable $\tau_M(S)$ corresponding to the minimal value of the diffusion coefficient $D$, i.e., the shift of the critical temperature:
\begin{enumerate}[label=\arabic*)]
	\item has a negative value in agreement with the scaling theory for finite-size systems~\cite{6,7},
	\item tends to zero with an increase of the geometric factor $S$ (linear size $L$),
	\item increases with transition from plane-parallel geometry to cylindrical geometry, i.e., with a decrease of the lower crossover dimensionality~\cite{13}.
\end{enumerate}

\section{The 2D--3D dimensional crossover for critical exponents and for spatial and fractal dimensions}
\label{sec-4}

In this section, another type of dimensional crossover is studied, namely DC-2, in which a smooth transition occurs with a change in critical exponents, as well as spatial and fractal dimensions with a decrease of the number $S$ of monomolecular layers. To be more precise, let us first consider a 2D--3D dimensional crossover for effective critical exponents, presented in table~\ref{tbl-smp3}. Note that the term ``effective critical exponents'' is used here for critical exponents that change their values with layer-by-layer ordering and depend on the number $S$ of monomolecular layers for the spatially confined systems under study.

{\itshape{\bfseries The 2D--3D dimensional crossover for critical exponents.}} An idea from the theory of mode coupling was used to obtain the following formula for any effective critical exponents  $n$ providing a smooth transition from its 3D $n_3$ to 2D $n_2$ values:
\begin{equation}
n=n_3+ \bigg[\frac{2}{\piup} \arctan (ax-b)-1\bigg]\frac{n_3-n_2}{2}.
\label{eq-23}
\end{equation} 	
Here, $x=L/L_0$  is the dimensionless width of the slitlike pore or radius of the cylindrical pore; $L_0$ is the linear size of the system at a reduced geometry where the dimensional crossover occurs; $a$ and $b$  are the dimensionless parameters characterizing the slope and position of the  $3D\leftrightarrow 2D$ crossover~\cite{11, 13}. The term ``effective critical exponents'' is used here in the sense that in layer-by-layer ordering, the critical exponents depend not only on the spatial dimension and on the order parameter dimension, i.e., on the number of independent components of the order parameter (see table~\ref{tbl-smp2}), but also on the number $S$ of monomolecular layers. It is important to note that the known scaling and hyperscaling (depending on the spatial dimension) equalities must be satisfied for all effective critical exponents at each fixed value of $S$.

\begin{figure}[h!]
\centerline{\includegraphics[width=0.65\textwidth]{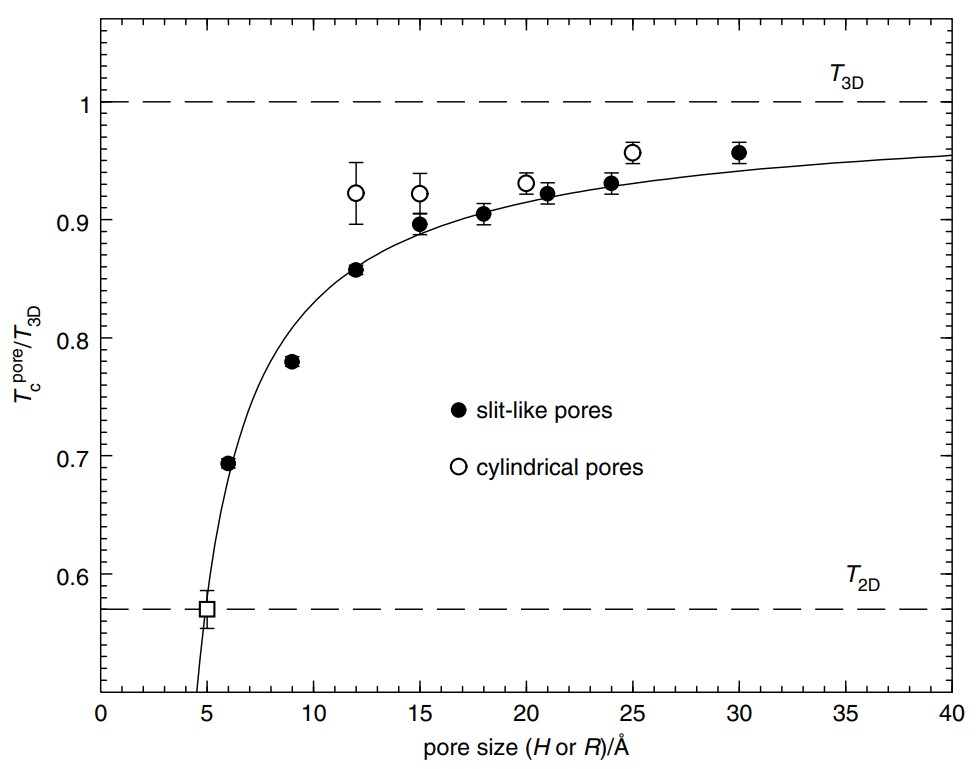}}
\caption{Size dependence of the pore critical temperature (computer experiment~\cite{11}).} \label{fig-1}
\end{figure}

\begin{figure}[h!]
\centerline{\includegraphics[width=0.65\textwidth]{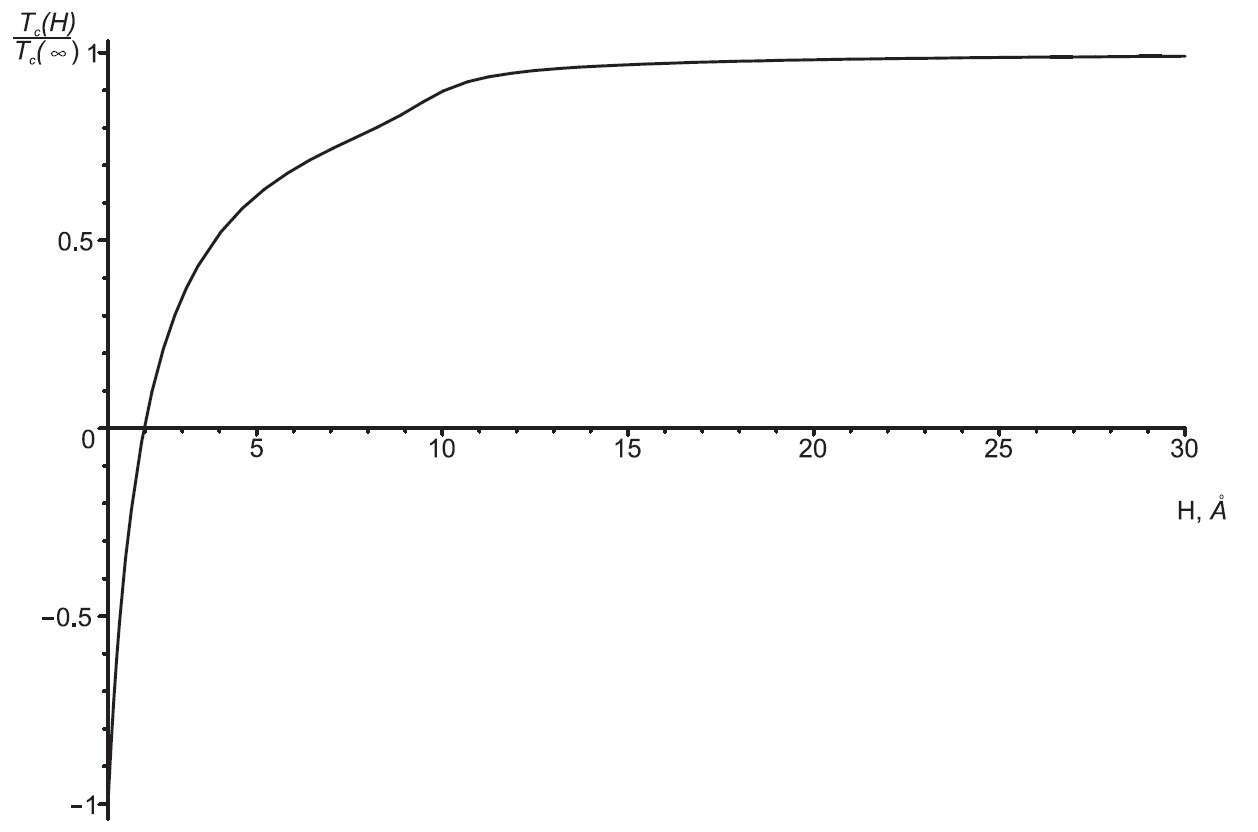}}
\caption{Size dependence of the critical temperature in a slitlike pore.} \label{fig-2}
\end{figure}

 Figures~\ref{fig-1},~\ref{fig-2} demonstrate the agreement between the data of ``computer experiment'' and our theoretical calculations for the dependence of the critical temperature $T_c(H)/T_c$ on the thickness $H$ of a slitlike pore filled with water, where $T_c(H)$ and $T_c$ are critical temperatures of confined and bulk water, respectively. Here, black circles in figure~\ref{fig-1} are the results of computer simulations~\cite{11}, solid line in figure~\ref{fig-2}  is the theoretical result obtained with the respect to  equation~(23)~\cite{1}.

     Let us consider a 2D--3D dimensional crossover of the 2nd type for the effective critical exponents, which are presented in table~\ref{tbl-smp3}. In the case of 2D systems with a geometry of the monomolecular layer or for the 2-dimensional Ising model, the critical exponents $\alpha = 0$,  $\beta= 1/8$,  $\gamma = 7/4$,   $\delta= 15$, $\nu  = 1$, while the critical exponents for 3D systems are as follows~\cite{35,36,37}: 
$\alpha = 0.110$, $\beta=0.3265$, $\delta=4.789$, $\gamma=1.237$, $\nu=0.630$.

\begin{table}[htb]
\caption{The effective critical exponents at 2D--3D dimensional crossover.}\vskip3mm\tabcolsep4.2pt
\label{tbl-smp3}
\centering{\begin{tabular}{|c|c|c|c|c|c|}
 \hline
 \multicolumn{1}{|c}
{\rule{0pt}{5mm} {\bfseries $S$}}
 & \multicolumn{1}{|c}{{\bfseries $\alpha_{\rm eff}$}}
& \multicolumn{1}{|c}{{\bfseries $\beta_{\rm eff}$}}
& \multicolumn{1}{|c}{{\bfseries $\delta_{\rm eff}$}}
& \multicolumn{1}{|c}{{\bfseries $\gamma_{\rm eff}$}}
& \multicolumn{1}{|c|}{{\bfseries $\nu_{\rm eff}$}}
\\%
\hline%
\rule{0pt}{5mm}
1&$\rightarrow$ 0 &$\rightarrow$ 0.125 &$ \rightarrow$ 15 & $\rightarrow$ 1.750 &$\rightarrow$ 1.000 \\%
\hline
2&0.025 &0.171 &10.544 & 1.632 &0.915 \\%
\hline
3&0.026 &0.173 & 10.416&1.629 &0.912\\%
\hline
4&0.027 &0.174 & 10.339&1.625 &0.910 \\%
\hline
5&0.028 &0.176 & 10.199&1.619 &0.906 \\%
\hline
6&0.030 &0.179 & 10.000&1.611 &0.900 \\%
\hline
7&0.032 &0.184 & 9.696&1.600 &0.892 \\%
\hline
8&0.036 &0.191 & 9.277&1.581 &0.878 \\%
\hline
9&0.044 &0.205 & 8.546&1.547 &0.853 \\%
\hline
10&0.059 &0.233 & 7.335&1.476 &0.802 \\%
\hline
11&0.084 &0.278 & 5.892&1.360 &0.719  \\%
\hline
12&0.102 &0.313 & 5.064&1.272 &0.655 \\%
\hline
13&$\rightarrow$ 0.110 &$\rightarrow$ 0.3265 & $\rightarrow$ 4.789& $\rightarrow$ 1.237 &$\rightarrow$ 0.630 \\%
\hline
\end{tabular}}
\end{table}

It should be noted that a  2D--3D dimensional crossover occurs in confined liquids if the number of molecular layers $S\approx 8$. Actually, Brovchenko and Oleinikova~\cite{11},  using the results of their computer simulations, showed that the critical index $\nu$  of the  temperature dependence of the correlation length changes from its bulk value $\nu = 0.63$ to the two-dimensional value $\nu = 1.0$  for water in  slitlike pores at a pore thickness about 2.4 nm, which corresponds to approximately 8 molecular diameters of water molecules.

{\itshape{\bfseries  The 2D--3D dimensional crossover for spatial and fractal dimensions.}} 
Let us introduce the Benoit Mandelbrot formula~\cite{39,40,41,42,43}
\begin{equation}
	D_f  =  d_{\rm eff}-\beta_{\rm eff}/\nu_{\rm eff},
	\label{eq-24}
\end{equation} 			 
considering this formula as a hyperscaling equality for the spatial dimension $d$ and a new critical index of the fractal dimension $D_f$. Table~\ref{tbl-smp4} contains the effective spatial $d_{\rm eff}$ and fractal $D_f$ dimensions calculated from a hyperscaling equality $d_{\rm eff} = (2-\alpha_{\rm eff})/\nu_{\rm eff}$   and the Mandelbrot's formula $d_{fr} = d_{\rm eff}-\beta_{\rm eff}/\nu_{\rm eff}$, taking into account  the interpolation formula (23) as well.

\begin{table}[h!]
\noindent\caption{The effective spatial $d_{\rm eff}$ and fractal $D_f$ dimensions at 2D--3D dimensional crossover. }\vskip3mm\tabcolsep4.2pt
\label{tbl-smp4}
\centering\noindent{\begin{tabular}{|c|c|c|c|c|c|}
 \hline
 \multicolumn{1}{|p{0.2\linewidth}}
{\rule{0pt}{5mm} {\small\bfseries The number $S$ of monomolecular layers}}
 & \multicolumn{1}{|p{0.3\linewidth}}{{\small\bfseries The effective spatial dimension $d_{\rm eff} = (2-\alpha_{\rm eff})/\nu_{\rm eff}$}}
& \multicolumn{1}{|p{0.3\linewidth}|}{{\small\bfseries The effective fractal dimension $D_f  =  d_{\rm eff}-\beta_{\rm eff}/\nu_{\rm eff}$}}

\\%
\hline%
\rule{0pt}{5mm}
1&2.000 & 1.875  \\%
\hline
2&2.158 & 1.971 \\%
\hline
3&2.163 & 1.973\\%
\hline
4&2.168 & 1.977 \\%
\hline
5&2.177 & 1.982 \\%
\hline
6&2.189 & 1.990 \\%
\hline
7&2.206 & 2.000 \\%
\hline
8&2.237 & 2.019 \\%
\hline
9&2.293 & 2.053 \\%
\hline
10&2.420& 2.129 \\%
\hline
11&2.665& 2.278 \\%
\hline
12&2.898& 2.742\\%
\hline
13&3.000& 2.482 \\%
\hline
\end{tabular}}
\end{table}

A $2D \leftrightarrow 3D$ DC-2 leads to the gradually varying dependence of $d_{\rm eff}$ and $D_f$ on the number $S$, which determines a fixed number of monolayers in the direction of the confinement of the system.

\section{Molecular light scattering and its biomedical application}
\label{sec-5}

Here, we are going to pay a special attention to such characteristics of the light critical opalescence spectrum in confined liquids as:  1) the width $\Gamma_{\rm MB}$ of Mandelstam--Brillouin components, 2) the frequency shift $\Gamma_{\rm MB}$ of the side (Mandelstam--Brillouin) components from the central (Rayleigh) line, 3) the Landau--Placzek relation $I_c/2I_{\rm MB}$~\cite{15,21,22,38}.

{\itshape{\bfseries The width of Mandelstam--Brillouin components.}} A leading contribution to the width $\Gamma_{\rm MB}$ of the  Mandelstam--Brillouin components is given by the following formula:
\begin{equation}
	\Gamma_{\rm MB}=\left\{ \left[ \frac{4}{3} \eta(q)+\zeta(q) \right] + \frac{\kappa(q)}{C_V(q)} -\frac {\kappa(q)}{C_p(q)}\right\}\frac{q^2}{\rho}.
	\label{eq-25}
\end{equation} 	
Here, $\eta$ and $\zeta$ are shear and bulk viscosities, $\kappa$ is the thermal conductivity, $C_p$ and $C_V$ are the specific heats at a constant pressure and volume, $\rho$ is the liquid density, $q=(2\piup/\lambda)\sqrt{2\varepsilon_0}(1-\cos\theta)^{1/2}$ is the change of wave vector on scattering by the angle $\theta$, $\lambda$ is the light wavelength, $\varepsilon_0$ is the average part of the dielectric permittivity.

Thus, the width $\Gamma_{\rm MB}$ of the Mandelstam--Brillouin components of the light critical opalescence spectrum satisfies the following formula for confined liquids (including liquids of biological nature):
\begin{enumerate}[label=(\alph*)]
	\item for a cylindrical geometry (e.g., ion channels, pores, etc.)
	\begin{equation}
		\Gamma_{\rm MB}\propto {\Gamma_{\rm MB}}^{(0)} \left[\tau+{\left(\frac{\mu_1\xi_0}{R}\right)}^{1/\nu} (1+\tau)\right]^{-3\nu};
		\label{eq-26}
	\end{equation} 	
	\item for a plane-parallel geometry (synaptic clefts, clefts, biological membranes, etc.)
	\begin{equation}
		\Gamma_{\rm MB}\propto {\Gamma_{\rm MB}}^{(0)} \left[\tau+{\left(\frac{\piup\xi_0}{H}\right)}^{1/\nu} (1+\tau)\right]^{-3\nu},
		\label{eq-27}
	\end{equation} 	
\end{enumerate}
where ${\Gamma_{\rm MB}}^{(0)}\cong \zeta_0 q^2/\rho$, $\zeta_0$ is the amplitude of the singular part of the bulk viscosity, $\xi_0$ is the amplitude of the correlation length, $\mu_1$ is the first zero of the Bessel function,  $\tau=(T_c - T)/T_c$ is the deviation of temperature, $\nu$ is critical exponent.

The width $\Gamma_{\rm MB}$ in confined liquids has no singularity at the critical temperature of a bulk liquid system $(\tau=0)$ and depends only on a geometric factor $L$ according to the relations

$$\Gamma_{\rm MB}\propto R^3$$ for a cylindrical geometry;

\begin{equation}\Gamma_{\rm MB}\propto H^3 	\label{eq-28}
\end{equation} 	 
for a plane-parallel geometry.   

{\itshape{\bfseries The frequency shift of Mandelstam--Brillouin components.}} The frequency shift $\Delta \Omega_{\rm MB}$ of the side (Mandelstam--Brillouin) components from the central (Rayleigh) line  is governed by the velocity of sound~$v(q)$ and in the hydrodynamic theory can be represented by the formula
\begin{equation}\Delta \Omega_{\rm MB} = v_{[\tau(L,q)]}q^2 . 	\label{eq-29}
\end{equation} 	
 The regular part of $\Delta \Omega_{\rm MB}$  is connected with the background part of the specific heat $C_V$ at the constant volume due to the relation $v_{\rm reg}\propto {(C_V)_{\rm reg}}^{-1/2}$, while the singular part of the frequency shift of $\Delta \Omega_{\rm MB}$ at low frequencies can be written as follows:
\begin{equation}
	(\Delta \Omega_{\rm MB})_S = v_S[\tau(L,q)]q^2\propto {(C_V)_{S}}^{-1/2} \propto {\tau(L,q)}^{\alpha/2}.	\label{eq-30}
\end{equation} 	
The main consequences obtained from equation~\eqref{eq-30}:
\begin{enumerate}[label=\arabic*)]
	\item A decrease in the frequency shift $\Delta \Omega_{\rm MB}$ is small because a numerical value $\alpha/2 \approx 0.05$ for classical liquids.
	\item The minimal value of $\Delta \Omega_{\rm MB}$ does not take place at the bulk critical temperature $T_c(\infty)$ but at the new ``critical temperature'' $T_c^*(L)$, as in the case of the width $\Gamma_{\rm MB}$.
	\item The frequency shift $\Delta \Omega_{\rm MB}$ in confined liquids has  no singularity at the critical temperature $T_c(\infty)$ of a bulk liquid system and depends on a geometric factor $L$:
	$$\Delta \Omega_{\rm MB} \propto  R^{-\alpha/2\nu} $$  
	for a cylindrical geometry, $R$ is the radius of a cylindrical pore;
	\begin{equation}\Delta \Omega_{\rm MB} \propto  H^{-\alpha/2\nu}  	\label{eq-31}
	\end{equation} 	 for a plane-parallel geometry, $H$ is the thickness of a slitlike pore.
\end{enumerate}

Equation~\eqref{eq-31} gives such a result: with decreasing a geometric factor $L$ of confined liquids (correspondinly, the thickness $H$ of a liquid film (e.g., biologic membranes, synaptic cleft, etc.) or the radius $R$ of cylindrical samples (e.g., ion channels, pores, etc.), the frequency shift of the side (Mandelstam--Brillouin) components of the light critical opalescence spectrum is weakly increasing: $\Delta \Omega_{\rm MB} \propto  L^{-\alpha/2\nu}\propto  L^{-0.087}$.

It is worth to mention that the results obtained above for the width $\Gamma_{\rm MB}$ and frequency shift $\Delta \Omega_{\rm MB}$ are valid only at frequencies $\omega \ll \omega_r$ where $\omega_r = \kappa / \rho C_V\xi^2 \propto \xi(L)^{-(1+\alpha/\nu)}$ is the relaxation frequency which decreases with approaching the new ``critical temperature'' ${T_c}^*(L)$ of confined liquids.

{\itshape{\bfseries The Landau--Placzek relation.}} As is well-known, the ratio $I_c/2I_{\rm MB}$ of the integral intensities of the central (Rayleigh) and side (Mandelstam--Brillouin) components is given by the  Landau--Placzek relation $I_c/2I_{\rm MB}=(C_P-C_V)/C_V$. While studying the light critical opalescence spectrum for confined liquids, one has the following formula for the Landau--Placzek relation in the hydrodynamic approximation:
\begin{equation}
	I_c/2I_{\rm MB} \propto \tau(L)^{-\gamma+\alpha}, 	\label{eq-32}
\end{equation} 	
with $-\gamma+\alpha \approx -1.1$. Thus, the Landau--Placzek relation~\eqref{eq-32} demonsrates a rapid growth of the integral intensity $I_c$ of the central component which becomes finite at the new ``critical temperature'' due to a spatial dispersion of the order-parameter fluctuation.

In conclusion of this section, the following remarks can be formulated as biomedical applications of the obtained results. With {\itshape decreasing (increasing)} the characteristic size $L$ of the system at a bulk critical temperature:
\begin{enumerate}[label=\arabic*)]
	\item the width $\Gamma_{\rm MB}$ of  Mandelstam--Brillouin components  strongly shortens (broadens) according to the relation $\Gamma_{\rm MB}\propto L^3$ [see equation~\eqref{eq-28}],
	\item the frequency shift $\Delta \Omega_{\rm MB}$ of the side  (Mandelstam--Brillouin) components from the central (Rayleigh) line  weakly increses (decreases) in accordance with such a relation $\Delta \Omega_{\rm MB} \propto  L^{-0.087}$ [see equation~\eqref{eq-31}],
	\item the Landau--Placzek relation essentially decreases (increases) as is seen from $I_c/2I_{\rm MB} \propto L^{1.79}$ [see equation~\eqref{eq-32}].
\end{enumerate}

The following important consequences can be obtained using the experimental data of the light critical opalescence spectra from suspension of tumor cells:
in case the membrane thickness $H$  increases 1.5~times (i.e., by 50 percent) during the process of proliferation, one should expect that
\begin{enumerate}[label=\arabic*)]
	\item the width of the central (Rayleigh) line shortens more than twice (2.25 times),
	\item the frequency shift $\Delta \Omega_{\rm MB}$ of the side  components decreases by 3.6  percent,
	\item the width of the side  (Mandelstam--Brillouin) components broadens more than 3 times (3.375~times).
\end{enumerate}

\section{Conclusions}
\label{sec-6}

The following conclusions are proposed as the results of reviewing and studying the effects of DC on physical properties of condensed systems near the phase transition and critical points.

	The basic theoretical concepts for describing the effects associated with the influence of DC on phase transitions and critical phenomena in condensed systems  were proposed and discussed. In particular, the following problems were consided: (a) correlation functions and the correlation length (radius) of the order parameter fluctuations in spatially limited liquids; (b) effective description of the dependence of equilibrium and kinetic properties on thermodynamic (temperature, density, pressure) and dimension variables for bulk and bounded liquids in the critical region; (c) inequalities describing the 3D transition from the dependence of physical properties on the thermodynamic variables to the description of these properties on the linear sizes of bounded systems. 

	The study of crossover phenomena in condensed systems with the Ginzburg number $\mathrm{Gi} <1$, and further studies of these crossover phenomena (including the DC-1 effects) were carried out using the diffusion coefficient of a two-component liquid mixture. The effects of the DC-2 on the dependence of the critical exponents of scaling power laws were studied with taking into account the corresponding interpolation formula for critical exponents and the number $S$ of molecular layers in the direction of spatial confinement at 3D--2D DC. 

	In addition, the consequences of including an equality for a critical index among the scaling and hyperscaling (containing spatial dimension $d$) equalities were investigated, being the Benoit Mandelbrot's formula and the fractal dimension $D_f$, respectively. 

	The results obtained from the application of the MLS method can be used for early diagnostics of malignant diseases, since the process of uncontrolled proliferation (tissue growth due to the cell division and enlargement) has a significant effect on the size dependences of the fine structure characteristics of the Rayleigh line.

\ukrainianpart

\title{Вплив розмірного кросовера на фазові переходи та критичні явища в конденсованих системах}
\author{О. В. Чалий , О. В. Зайцева  }
\address{
	Національний медичний університет імені О. О. Богомольця, бульвар Тараса Шевченка 13, 01601 Київ, Україна
}

\makeukrtitle

\begin{abstract} 
	Ця стаття спрямована на вивчення впливу розмірного кросовера (РК) на фізичні властивості конденсованих систем поблизу фазового переходу та критичних точок. У статті розглянуті такі проблеми: (1) теоретичні обґрунтування, що дозволяють вивчати вплив просторового обмеження на РК поблизу фазового переходу та критичних точок; (2) дослідження РК у конденсованих системах з числом Гінзбурга $\mathrm{Gi} <1$, де флуктуаційні ефекти описуються по-різному у флуктуаційній, регулярній та проміжній (кросоверній) областях; (3) дослідження двох типів РК, коли (а) флуктуації призводять до перетворення залежності від термодинамічної змінної на залежність від лінійних розмірів 3D-систем зі зменшенням лінійних розмірів $L$ до значень радіусу кореляції параметра порядку, а також (б) з подальшим зменшенням лінійних розмірів $L$ виникає 3D--2D або 3D--1D РК залежно від геометрії (щілиноподібної або циліндричної), яка визначається значенням нижньо-кросоверної розмірності $d_{\rm LCD}$; (4) пропонується розширити відомі рівності для критичних індексів, використовуючи формулу Мандельброта для фрактальної розмірності $D_f$ як критичний показник; (5) досліджується вплив РК 3D--2D на характеристики тонкої структури спектра молекулярного розсіяння світла (МРС).
	
	\keywords 
	розмірний кросовер, критичні індекси, обмежені (замкнені) рідини, нижня кросоверна розмірність, фрактальна розмірність, коефіцієнт дифузії 
\end{abstract}

\lastpage
\end{document}